\documentclass[twocolumn,  superscriptaddress, pra]{revtex4}
\usepackage{amsmath}
\usepackage{amssymb}
\usepackage{graphicx}
\usepackage{booktabs}
\usepackage{subfigure}
\usepackage{esint}
\usepackage{tabularx}

\begin{document}

\title{Powerful harmonic charging in quantum battery}
\author{Yu-Yu Zhang}
\email{yuyuzh@cqu.edu.cn}
\affiliation{Department of Physics, Chongqing University, Chongqing 401330, China}
\affiliation{Graduate School, China Academy of Engineering Physics, Beijing 100193, China}
\author{Tian-Ran Yang}
\affiliation{Department of Physics, Chongqing University, Chongqing 401330, China}
\author{Libin Fu}
\email{lbfu@gscaep.ac.cn}
\affiliation{Graduate School, China Academy of Engineering Physics, Beijing 100193, China}
\author{Xiaoguang Wang}
\email{xgwang1208@zju.edu.cn}
\affiliation{Zhejiang Institute of Modern Physics, Department of Physics, Zhejiang
University, Hangzhou 310027, China}
\date{\today}

\begin{abstract}
We consider a harmonic charging field as an
energy charger for the quantum battery, which consists of an ensemble of
two-level atoms. The charging of noninteracting atoms is completely
fulfilled, which exhibits a substantial improvement over previous static charging
fields. Involving the repulsive interactions of atoms, the fully
charging is achieved with shorter charged period over the noninteracting
case, yielding an advantage for the charing. Excluding the charging field, a
quantum phase transition is induced by the attractive atom-atom
interactions, and the interacting atoms become to be degenerate in
the ground state. We find that the degenerate states play a
negative role in the charging due to the gapless energies.
The atoms with strong attractive interactions can not be charged completely,
which is accompanied by a drop of the maximum stored energy.
\end{abstract}

\maketitle

\section{Introduction}

Quantum information science develops very quickly in recent years. Various
kinds of tasks using quantum information have been studied in detail such as
quantum sensing, computations, and communications. Among these proposals,
quantum battery (QB) was proposed to use quantum effects such as quantum
correlations to enhance the charging power and speed up the charging time in
comparison with its classical counterpart~\cite%
{Hovhannisyan2013,Friis2016,Friis17,Farina18}. The concept of a QB was
originally proposed as a two-level system used to temporarily store energy
transferred from an external field\cite{Alicki2013,Binder2015}. How to make
an efficient energy storage by exploiting nonclassical effects is central
and practical research subject.

Recent research efforts have been devoted to explore contributions provided
by quantum correlations for charging in collective QBs~\cite%
{Binder2015,Campaioli2017,Le2018,Ferraro2018,Andolina2018_1,fusco2016}. The
Dicke QBs~\cite{Ferraro2018,fusco2016} describe collective QBs coupling to
one common cavity, which serves as a global charger. The corresponding
QB-charger coupling produces indirect interactions between QBs.
Consequently, the advantage of quantum correlations of QBs mediated by the
global charger has been explored in the charging power of the Dicke QBs. By
contrast, there is another kind of QBs, in which the quantum correlations
are induced by intrinsic interaction between QBs~\cite{Le2018}. There is a
physical phenomenon that shares many of the features with the quantum
correlations in interacting systems---the quantum phase transition, which is
induced by the change of a coupling parameter. It is interesting to study
the properties of QBs in different phases related to quantum correlations,
which should be on a close relation to the collective charging. For example,
the connection between the phase transition induced by the QB-charger
coupling and the optimal energy storage has recently been studied~\cite%
{Ferraro2018,fusco2016}. However, the effects of quantum correlations
in the other kind of QBs with intrinsic
interactions between batteries, especially a phase transition, are overlooked.
On the other hand, in all
these studies QBs were investigated with a static charging field~%
\cite{Alicki2013,Binder2015,Le2018,fusco2016,Compaioli2018,Campaioli2017}.
Although a harmonic driving as a charger has been studied numerically~\cite%
{Farina18,Ferraro2018,Andolina2018_1,Andolina2018_2}, an analytical
solution for the harmonic charging remains elusive. It is
challenge to solve a time-dependent Hamiltonian analytically to give an
optimal driving frequency for the charging of the QB.

In this paper, we consider a QB system of $N$ two-level atoms
and a semiclassical harmonic field as a charger.
The quantum correlations of the atoms rely on the interatomic infinite-range interactions.
The charging of the QB with non-interacting atoms can be completely fulfilled,
and the optimal driving frequency of the harmonic charger is obtained analytically.
It exhibits a substantial improvement over the
previous static charging field. Involving repulsive intrinsic interactions
of atoms, the QB can be charged faster than that of noninteracting
atoms. While the optimal charging period for the attractive interactions becomes longer as the
coupling strength increases. For strong attractive interactions
there occurs a quantum phase transition, which is accompanied by degenerate
energies of the ground state. We find the maximal
stored energy of the atoms in the degenerate phase drops from the fully-charged value, which
indicates that the gapless states plays a negative role in the charging process.

The paper is outlined as follows. In Sec. II, we study $N$ two-level atoms
charging independently by the harmonic field. The maximum stored energy and
the optimal driving frequency are given analytically. In Sec. III, we
discuss contributions of the repulsive and attractive interatomic interactions
in the process of energy storage.
Finally, a brief summary is given in Sec.~IV.
\begin{figure}[tbp]
\includegraphics[scale=0.25]{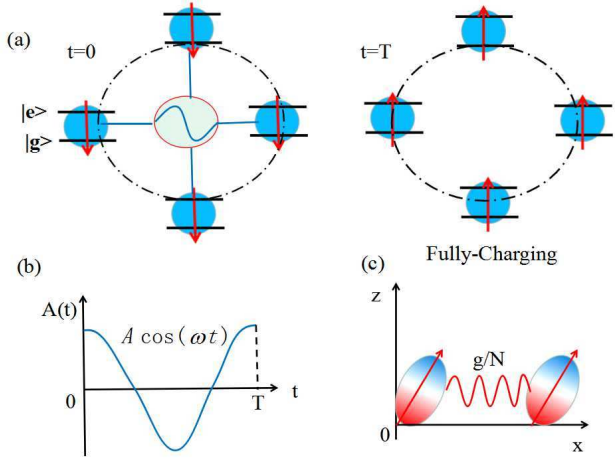}
\caption{(a) Charging protocol of $N$ two-level atoms as the QB. At time $%
t=0 $ each atom is in the ground state $|g\rangle$. At the
period $T$, the QB is fully charged and the final state of atoms is $|e\rangle^{\otimes N}$%
. (b)The QB is charged with a harmonic driving field $%
A\cos(\protect\omega t)$. During the charging time $0<t<T$, the QB
interacts with a harmonic driving field $A\cos(\protect\omega %
t)$. Finally, the interaction is switched off at the end of charging period
$T$. (c) The intrinsic infinite-range interactions between arbitrary two atoms include the
repulsive ($g>0$) and attractive ($g<0$) coupling. }
\label{charging}
\end{figure}

\section{ Charging for noninteracting atoms}

The QB consists of an ensemble of independent two-level atoms, which are
collective charged by a harmonic field in Fig.~\ref{charging}(a). The
Hamiltonian of $N$ noninteracting atoms is given as
\begin{equation}
H_{0}=\frac{\Delta }{2}\sum_{i=1}^{N}\sigma _{i}^{z}=\Delta S_{z},
\end{equation}
where the collective atom operators $S_{\alpha }=\sum_{i}\sigma _{i}^{\alpha
}/2$ ($\alpha =x,y,z$), $\Delta $ is the energy level splitting of the
two-level atom. The basis set for representing the atoms system is the Dicke
states $|S,m\rangle $ ($m=-S,-S+1,...,S$), which are eigenstates of $S^{2}$
and $S_{z}$ with the total pseudospin $S=N/2$. We set $\Delta =1$ in the
following.

We employ a harmonic field as a charger to transfer energy to the battery as
much as possible. All two-level atoms are driven by the harmonic charging
field as
\begin{equation}  \label{driving}
H_{1}=\frac{A}{2}\cos(\omega t)\sum_{i=1}^{N}\sigma _{i}^{x}=A\cos(\omega
t)S_{x},
\end{equation}%
where $A$ and $\omega $ are the driving amplitude and the modulated
frequency. For a comparison, the driving Hamiltonian with a static charging
field is $H_{1,s}=AS_{x}$. It is noted that the coupling between QB and the
harmonic charger is the conchoidal function of time instead of a constant in
the static charger. Fig.~\ref{charging}(b) shows the charging procedure is
designed to turn on the interaction between $N$ atoms and the harmonic field
during the charging interval $0<t<T$. Then the interaction is turned off at
time $T$, and the QB is isolated from the external field and keep their
energy. The charging period is associated with the alternative driving
frequency as $T=2\pi/\omega$. During the charging step, the total
Hamiltonian for $N$ two-level atoms interacting with the harmonic field is $%
H=H_{0}+H_{1}$, which is viewed as the collective Hamiltonian.

To study the advantage for the charging with the harmonic charging
field, we focus on maximizing the stored energy in the QB and minimizing the
charging time. Initially, $N$ two-level atoms are prepared in the
lowest-energy state as $|\varphi _{N}(0)\rangle =|N/2,-N/2\rangle $, for which
each atom is in the ground state $|g\rangle $. The
wave function evolutes according to the schr\"{o}dinger equation $%
i\partial|\varphi _{N}(t)\rangle/\partial t =H|\varphi _{N}(t)\rangle $. At
the end of charging period $T$, the stored energy that moves from the
harmonic field to the QB can be expressed in terms of the mean local energy
of the QB, i.e.~\cite{Ferraro2018,Le2018}
\begin{equation}  \label{stored energy1}
E_{N}(T)=\langle \varphi _{N}(T)|H_{0}|\varphi _{N}(T)\rangle -\langle
\varphi _{N}(0)|H_{0}|\varphi _{N}(0)\rangle.
\end{equation}

A battery is a physical system that stored energy in atoms, which is
transferred from the charging field. We investigate the maximum stored
energy $E_{N,\max }$ during the charging process. The advantage of the
harmonic field lies in the modulated frequency $\omega $, which can be tuned
to produce maximum stored energy $E_{N,\max }$ at an optimal charging period
$T_{\max}=2\pi /\omega_{\max} $. As the QB is charged completely, each of
two-level atoms is in the upper state and the scaled stored energy $E_{\max
}(T)/N\Delta $ is expected to be the fully-charging value $1$. The
corresponding final state at the end of the charging is
\begin{equation}  \label{psi1}
|\varphi _{N}(T_{\max })\rangle =|e\rangle ^{\otimes N},
\end{equation}
which is called the fully-charging state in Fig.1 (a).

Inspired by the approximated analytical solution for the driven
semi-classical Rabi model for a driving two-level system~\cite{zheng15}, we
extend the approach to solve the dynamics of $N$ two-level atoms
analytically. Using a unitary transformation $U=\exp [i\frac{A}{\omega \sqrt{%
N}}\xi \sin (\omega t)S_{x}]$ with the undetermined parameter $\xi \in
\lbrack 0,1]$, one obtain $H^{\prime }=UHU^{\dagger }-iU\frac{d}{dt}%
U^{\dagger }$ as
\begin{eqnarray}
H^{\prime } &=&\Delta \big\{\cos \big[\frac{A}{\omega \sqrt{N}}\xi \sin
(\omega t)\big]S_{z}+\sin \big[\frac{A}{\omega \sqrt{N}}\xi \sin (\omega t)%
\big]S_{y}\big \}  \notag \\
&&+A(1-\frac{\xi }{\sqrt{N}})\cos (\omega t)S_{x}.
\end{eqnarray}%
We expand the operator identities $\cos \big[\frac{A}{\omega \sqrt{N}}\xi
\sin (\omega t)\big]=J_{0}\big(\frac{A}{\omega \sqrt{N}}\xi \big)%
+2\sum_{n=1}^{\infty }J_{2n}\big(\frac{A}{\omega \sqrt{N}}\xi \big)\cos
(2n\omega t)$ and $\sin \big[\frac{A}{\omega \sqrt{N}}\xi \sin (\omega t)%
\big]=2\sum_{n=0}^{\infty }J_{2n+1}\big(\frac{A}{\omega \sqrt{N}}\xi \big)%
\sin [(2n+1)\omega t]$, where $J_{n}(\frac{A}{\omega \sqrt{N}}\xi )$ denotes
the Bessel function of integer order $n$. Then, we reasonably neglect all
higher-order harmonic terms ($n\geq 2$) with the higher-order Bessel
functions $J_{n}(\frac{A}{\omega \sqrt{N}}\xi )$. The Hamiltonian is
approximated as%
\begin{eqnarray}
H^{\prime } &=&\Delta J_{0}\big(\frac{A}{\omega \sqrt{N}}\xi \big)S_{z}+A(1-%
\frac{\xi }{\sqrt{N}})\cos (\omega t)S_{x}  \notag \\
&&+2\Delta J_{1}\big(\frac{A}{\omega \sqrt{N}}\xi \big)\sin (\omega t)S_{y}.
\end{eqnarray}%
The Hamiltonian includes the counter-rotating (CR) terms $e^{i\omega
t}S_{+}+e^{-i\omega t}S_{-}$ and the rotating-wave terms $e^{i\omega
t}S_{-}+e^{-i\omega t}S_{+}$. Since the CR terms describe fast oscillation
and virtual atom-field interacting processes, it is reasonable to make these terms vanish. We
can choose the parameter $\xi $ to tune the coefficient of the CR terms to
be zero, giving
\begin{equation}
A(1-\frac{\xi }{\sqrt{N}})-2\Delta J_{1}(\frac{A}{\omega \sqrt{N}}\xi )=0.
\label{xi}
\end{equation}%
Then $\xi $ is determined as $\bar{\xi}$.
Consequently, the transformed Hamiltonian becomes
\begin{equation}
H^{\prime \prime }=\Delta J_{0}\big(\frac{A}{\omega \sqrt{N}}\bar{\xi}\big)%
S_{z}+\tilde{A}(e^{i\omega t}S_{-}+e^{-i\omega t}S_{+}),  \label{H2}
\end{equation}%
where $\Delta J_{0}(\frac{A}{\omega \sqrt{N}}\bar{\xi})$ is the renormalized
atomic transition frequency, and $\tilde{A}=\frac{A}{2}(1-\frac{\bar{\xi}}{%
\sqrt{N}})$ is the effective coupling strength between the QB and the
charging field.

Using a unitary transformation $S=\exp (-i\omega tS_{z})$, the time-independent
Hamiltonian $\tilde{H}=SH^{\prime \prime }S^{\dagger}-iSdS^{\dagger}/dt$ is given by
\begin{equation}
\tilde{H}=\tilde{\Delta}S_{z}+2\tilde{A}S_{x},  \label{He}
\end{equation}%
where the effective detuning is
\begin{equation}
\tilde{\Delta}=\Delta J_{0}\big(\frac{A}{\omega \sqrt{N}}\bar{\xi}\big)%
-\omega .  \label{Deltap}
\end{equation}%
In the rotating frame, $\tilde{H}$ can be solved independent on the time.
By contrast to a static charging field, the
advantage of the harmonic charger lies in the renormalized detuning $\tilde{%
\Delta}$ and the effective QB-charger coupling strength $\tilde{A}$ of the
effective Hamiltonian $\tilde{H}$, which
can be tuned by the driving frequency $\omega $.

The charging of $N$ noninteracting atoms is equivalent to parallel charging
for independent atoms, and the scaled stored energy $E_{N}/(N)$ equals to $%
E_{1}$ of the single-atom battery. So we focus on the energy storage in the single atom.
For $N=1$, the effective Hamiltonian $\tilde{H}$ (\ref{He}) with $S=1/2$ can be solved analytically. The eigenvalues are
given as $\varepsilon _{\pm }=\pm \Omega _{R}/2$, where $\Omega _{R}$ is the
effective Rabi frequency
\begin{equation}
\Omega _{R}=\sqrt{\tilde{\Delta}^{2}+4\tilde{A}^{2}}.
\end{equation}
The corresponding eigenstates are dressed states $\mathtt{sin}\theta|\mp
z\rangle\pm \mathtt{cos}\theta|\pm z\rangle$ with $\mathtt{tan}(2\theta)=2\tilde{A}/\tilde{\Delta}$,
where $|\pm z\rangle$ are the eigenstates of $S_z$.

At the end of the charging protocol, the final state of the one-atom battery is
given explicitly by the eigenstates and eigenvalues as
\begin{eqnarray}  \label{psi}
|\varphi _{1}(T)\rangle &&=-i\frac{2\tilde{A}}{\Omega _{R}}\sin (\varepsilon
_{+}T)|e\rangle  \notag \\
&&+\big[\cos (\varepsilon _{+}T)+i\frac{\tilde{\Delta}}{\Omega _{R}}\sin
(\varepsilon _{+}T)\big]|g\rangle .
\end{eqnarray}

\begin{figure}[tbp]
\includegraphics[scale=0.5]{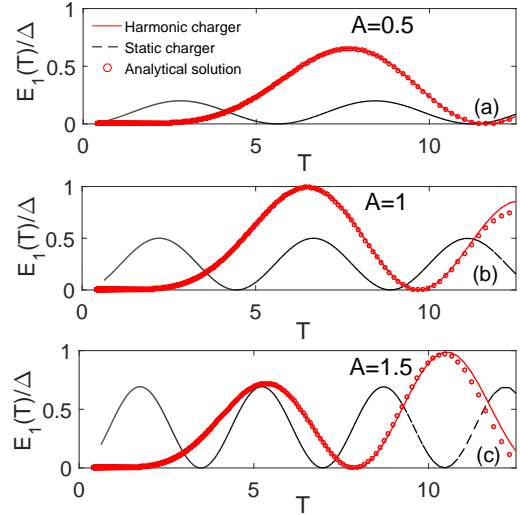}
\caption{Stored energy $E_1(T)/\Delta $ in the single-atom battery as a
function of charging period $T$ for driving amplitude $A=0.5$(a), $A=1$(b)
and $A=1.5$ (c). The QB
couples to the harmonic charger $A\cos(\protect\omega t)$ (red sold line)
and the static charger $A$ (black dashed line), respectively. The analytical
results of $E_1(T)/\Delta $ in Eq.(\ref{stored energy}) for the
harmonic charging filed are shown in red circle.}
\label{single}
\end{figure}

The corresponding stored energy in the single-atom battery is
\begin{equation}
E_{1}(T)/\Delta =\frac{2\tilde{A}^{2}}{\Omega _{R}^{2}}[1-\cos (\Omega
_{R}T)].  \label{stored energy}
\end{equation}
The analytical stored energy $E_1(T)/\Delta$ is consistent with numerical
results in a wide range of the charging period $T$ in Fig.~\ref{single}.

At the optimal period $T_{\max }=n\pi /\Omega _{R}$ ( for odd integer $%
n=1,3,...$), the maximal value of the stored energy in Eq.(\ref{stored energy}) is given as
\begin{equation}
E_{1,\max }/\Delta =\frac{4\tilde{A}^{2}}{\tilde{\Delta}^{2}+4\tilde{A}^{2}}.
\label{maximum}
\end{equation}%
The corresponding optimal driving frequency $\omega _{\max }$
is determined as $2\pi /T_{\max }=2\Omega _{R}/n$. Fig.~\ref{single} shows that the stored energy $E_{1}(T)/\Delta$
has local maximal values at a few peaks, which depends on the optimal charging period
$T_{\max}$ with the odd integer $n$. It is obvious that $E_{1,\max }/\Delta $
in Eq.(\ref{maximum}) ranges from $0$ to $1$ dependent on the effective transition frequency of
atoms $\tilde{\Delta}$ in Eq.(\ref{Deltap}), which is a function of the
driving frequency $\omega $. In particular, one can achieve the fully-charging value $E_{1,\max
}/\Delta =1$ by modulating $\omega $ to
satisfy $\tilde{\Delta}=0$. It leads to the fully-charging condition
\begin{equation}
\omega _{\max }=\Delta J_{0}(\frac{A}{\omega }\bar{\xi})  \label{omega}
\end{equation}%
with $\bar{\xi}$ determined in Eq. (\ref{xi}).
The final state $|\varphi _{1}(T)\rangle$ in Eq.(\ref{psi}) evolutes to be $|e\rangle$, which
demonstrate that each atom is completely charged.

\begin{figure}[tbp]
\includegraphics[scale=0.45]{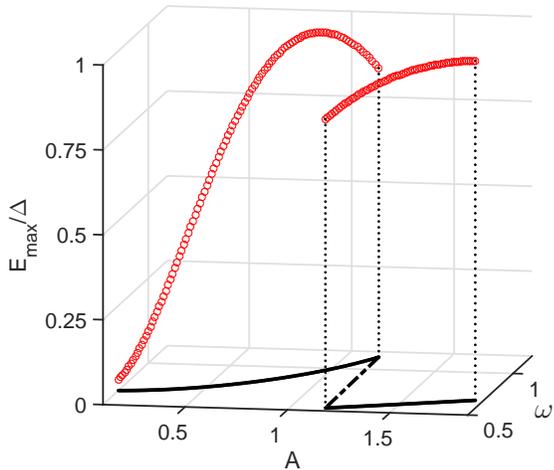}
\caption{Maximum stored energy $E_{\max}/\Delta $ (red circles) in the
single-atom battery as a function of the driving frequency $\protect\omega$
and the amplitude $A$ of the harmonic charger $A\cos(\protect\omega t)$. The
contour projection displays the optimal frequency $\protect\omega _{\max}$
(solid black line) dependent on $A$.}
\label{Aomega}
\end{figure}

Fig.~\ref{Aomega} displays the maximum stored energy $E_{1,\max }/\Delta $
for different driving amplitude $A$. The contour projection of $E_{1,\max
}/\Delta $ presents the optimal frequency $\omega_{\max} $. As $A$ increases
to $1$, $E_{1,\max }/\Delta $ increases to $1$ and then decreases.
The maximum stored energy has a jumps around $A=1.2$. It
ascribes to the discontinuous jump of the optimal frequency $%
\omega_{\max}$ with the odd integer $n$ changing from $1$ to $%
3$.

For a comparison, the Hamiltonian of the QB coupled with the static charging
field is $H_{s}=\Delta S_z+AS_{x}$. Similarly, at the end of charging time $%
T $ the energy stored in the battery is
\begin{equation}
E_{s}(T)/\Delta =\frac{1}{2}\frac{A^{2}}{\Delta ^{2}+A^{2}}[1-\cos(\sqrt{%
\Delta ^{2}+A^{2}}T)].
\end{equation}%
Obviously, the maximum stored energy is given by $E_{s,\max }/\Delta
=A^{2}/(\Delta ^{2}+A^{2})$. It is impossible to achieve the fully-charging
value $E_{s,\max }=1$. It means that the charging of the QB with a static
charging field is not completely fulfilled. The maximum stored energy with
the harmonic charging field is larger than that with the static charging
field in Fig.~\ref{single}, exhibiting the powerful harmonic charging.

\section{Collective charging with interatomic correlations}
With the consideration of additional interatomic interactions, a quite
natural question follows as to the effects on the charging battery. Quantum
correlations in multipartite systems is connected to energy storage~\cite%
{Alicki2013,Hovhannisyan2013,Binder2015,Campaioli2017}. It is interesting to
study the positive and negative effects of the quantum correlations induced
by intrinsic interactions of two-level atoms in the charging of the QB.

For $N$ identical two-level atoms, long-range forces between all atoms can
be mediated by the electric field. Such long-range interactions can be
engineered and controlled using atoms trapped in a photonic crystal
waveguide~\cite{kimble} and Bose-Einstein condensed atoms~\cite{baumann10},
which highlight the practical relevance for the interacting Hamiltonian
considered here. Each two-level atom is polarized in Fig.~\ref{charging}(c),
which can be described as an electric dipole operator $\hat{d}=d\sigma
_{+}+d^{\ast }\sigma _{-}$ with the dipole momentum $d$ and $d^{\ast }$.
Involving the infinite-range dipole-dipole interactions, the
Hamiltonian of $N$ interacting atoms can be described by~\cite{chen11,wang08}
\begin{eqnarray}  \label{h0interacting}
H_0^{I} &=&\frac{\Delta }{2}\sum_{i=1}^{N}\sigma _{i}^{z}+\frac{g}{2N}%
\sum_{i\neq j}^{N}(\sigma _{i}^{x}\sigma _{j}^{x}+\sigma _{i}^{y}\sigma
_{j}^{y})  \notag \\
&=&\Delta S_{z}+\frac{g}{N}(S^{2}-S_{z}^{2}-\frac{N}{2}),
\end{eqnarray}%
where $g$ is the atom-atom coupling strength including the repulsive ($g>0$)
and attractive $(g<0)$ interactions. We define the scaled coupling strength $%
\lambda=g/\Delta $.

\begin{figure}[tbp]
\includegraphics[scale=0.43]{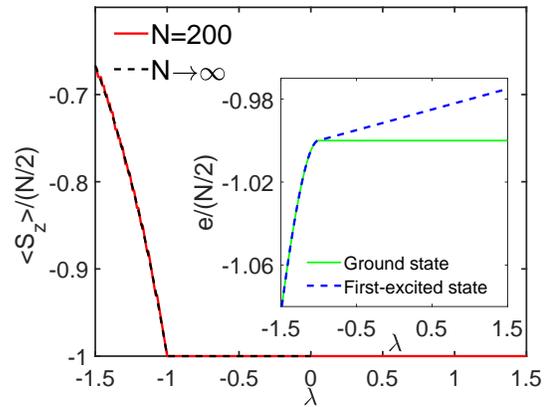}
\caption{$\langle S_{z}\rangle/(N/2)$ as a function of the repulsive and
attractive coupling strength $\protect\lambda$ for $N=200$ atoms. The
analytical results of $\langle S_{z}\rangle/(N/2)$ for the infinite atoms $%
N\rightarrow\infty$ displays the quantum phase transition at $\protect\lambda%
_c=-1$ in the attractive interactions case (black dashed line). The inset
shows the scaled energy $e/(N/2)$ for the ground state(green solid line) and the first-excited state
energy (blue dashed line) dependent on $\protect\lambda$.}
\label{phase1}
\end{figure}
The interacting Hamiltonian $H_0^{I}$ describes long-range interactions in
two-level systems such as Lipkin-Meshkov-Glick (LMG) model~\cite{LMG65,sun07}%
. The ground state of $H_0^{I}$ in Eq.(\ref{h0interacting}) lies in the
subspace spanned by the Dicke states $\{|N/2,M\rangle, M=-N/2,...N/2\}$ with
the total spin $S=N/2$, which is the eigenstate of $S_z$ with the eigenvalue
$M$. The Hamiltonian reduces into two spaces dependent on even or odd values
of $(N/2-M)$ and it is denoted as parity. The interacting atoms with
attractive interactions ($g<0$) undergo a quantum phase transition due to the competition
between the first noninteracting term and the second interacting terms of $H_0^{I}$ .
For a weak attractive coupling strength, the even and odd levels are
obviously separated. The ground state is fully polarized in the $Z$%
-direction and is given as $|N/2,-N/2\rangle$. When $\lambda$ exceeds
the critical attractive coupling strength, the even and odd levels
become degenerate in the thermodynamical limit. The inset of Fig.~\ref%
{phase1} displays that energy levels $e/(N/2)$ of $N=200$ atoms in the ground state and the
first-excited state are almost degenerate for $\lambda<-1$.
However, for arbitrary repulsive coupling strength ($g>0$), the ground state
remains $|N/2,-N/2\rangle$ with the lowest energy $-\Delta N/2$, and the
energy levels are non-degenerate.

\begin{figure}[tbp]
\includegraphics[scale=0.4]{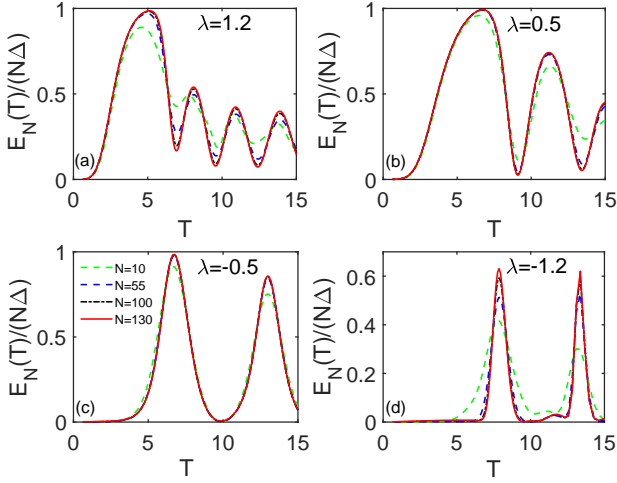}
\caption{Stored energy $E_N(T)$ (in unit of $N\Delta $) as a function of
charging period $T$ for different $N$ with interatomic coupling strength $%
\protect\lambda =1.2$(a), $\lambda=0.5$(b), $\lambda=-0.5$(c), and $\lambda=-1.2$(d).
The driving amplitude is $A=1$.}
\label{lambda}
\end{figure}
To explore the phase transition in the attractive interactions, we now use
the Hosltein-Primakoff transformation to study the infinite atoms in terms
of auxiliary bosonic operators $b^{\dagger }$ and $b$: $S_{z}=b^{\dagger
}b-N/2$ and $S_{+}=b^{\dagger }\sqrt{N}$. Above the critical coupling value $%
\lambda _{c}$, the bosonic field is expected to shift with a value $\beta $
as $b^{\dagger }\rightarrow b^{\dagger }+\beta $. We then obtain the
approximated Hamiltonian
\begin{eqnarray}
\tilde{H_0^{I}} &=&\Delta \lbrack (b^{\dagger }+\beta )(b+\beta )-\frac{N}{2}%
]  \notag \\
&&+\frac{g}{N}\big\{\frac{N^{2}}{4}-[(b^{\dagger }+\beta )(b+\beta )-\frac{N%
}{2}]^{2}\big\}.
\end{eqnarray}%
To make the linear terms ($b^{\dagger }+b$) vanish, one obtain%
\begin{equation}
\beta ^{2}=N(g+\Delta )/(2g).
\end{equation}%
Obviously, the critical value is given by $\lambda _{c}=g_{c}/\Delta=-1$.
Above the critical value, the expected value of atom polarization in the
ground state is given by $\langle S_{z}\rangle =\beta ^{2}-N/2$. Fig.~\ref%
{phase1} shows the behavior of the polarized value per atom $\langle
S_{z}\rangle /(N/2)$ for infinite atoms $N\rightarrow \infty $, exhibiting a
quantum phase transition at $\lambda _{c}=-1$.

It is interesting to study whether the attractive and repulsive interactions,
especially the phase transition, can enhance the charging of the QB.
The stored energy in the $N$ interacting atoms can be expressed as
\begin{equation}
E_{N}(T)=\langle \varphi _{N}(T)|H_0^{I}|\varphi _{N}(T)\rangle -\langle
\varphi _{N}(0)|H_{0}^I|\varphi _{N}(0)\rangle,
\end{equation}
where the state evolutes according to
$i\partial|\varphi_{N}(T)\rangle/\partial t=(H_0^{I}+H_1)|\varphi_{N}(T)\rangle$
with the charging harmonic field $H_1$ in Eq.(\ref{driving}).
Due to the intrinsic many-body interactions, the energy stored
in the interacting atoms is in
general a complicated function of the charging period $T$.

\begin{figure}[tbp]
\includegraphics[scale=0.45]{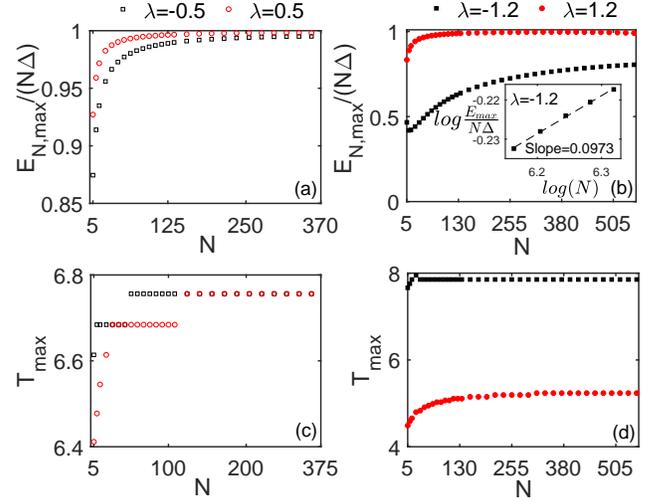}
\caption{Maximum stored energy in the battery $E_{N,\max }/(N\Delta) $
and the optimal charging period $T_{\max }$ as a function of
atoms number $N$ for interatomic coupling strength $\lambda=\pm 0.5$ (open circle)(a)(c) and $\protect%
\lambda =\pm 1.2$ (solid circle)(b)(d), respectively.
For $\lambda=-1.2$, the inset shows $E_{\max}/(N\Delta)$ versus $N$ on a log-log scale,
showing the slope of the scaling line $0.0973$(b). }
\label{et}
\end{figure}
We numerically calculate the stored energy $E_{N}(T)$ dependent on the
coupling strength $\lambda$ for the driving amplitude $A=1$. The dependence
of the rescaled stored energy $E_{N}/(N\Delta )$ on the charging period $T$
is shown in Fig.~\ref{lambda}. We observe the maximum stored energy $%
E_{N,\max}(T)\equiv \max_{T}E_{N}(T)$ locates at the first peak, where the
optimal charging period $T_{\max}$ is determined. The corresponding optimal
driving frequency is given by $\omega_{\max}=2\pi/T_{\max}$.

As the system size $N$ increases to $600$, $E_{N,\max}/(N\Delta)$ converges to be $1$ for the
repulsive coupling strength $\lambda=0.5$ and $1.2$ in Fig.~\ref{et}(a) and
(b). For the attractive interacting atoms with $\lambda=-1.2$,
$E_{\max}/(N\Delta)$ also converges with a lower value in Fig.~\ref{et}%
(b), for which the slope of the scaling line approaches zero.
It demonstrates that the scaling laws of the maximum stored energy for the repulsive and
attractive interacting atoms is
\begin{equation}
E_{N,\max} \propto N,
\end{equation}
which is the same as results of the Dicke quantum battery~\cite{Ferraro2018}.
On the other hand, the optimal charging period $T_{\max }$ converges to the same value
for weak attractive and repulsive coupling strength $\lambda=\pm0.5$ in Fig.~\ref{et}(c).
However, as the attractive interactions become strong with $\lambda=-1.2$, $T_{\max }$ becomes
longer than that for the repulsive coupling strength $\lambda=1.2$ in Fig.~\ref{et}(d).

\begin{figure}[tbp]
\includegraphics[scale=0.5]{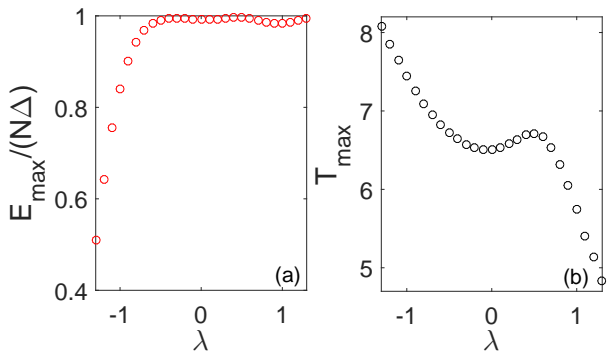}
\caption{(a)Maximum stored energy $E_{\max}$ (in unit of $N\Delta$) and (b) the
optimal charging period $T_{\max}$ dependent on the interatomic coupling
strength $\protect\lambda$ for $N=140$ atoms.}
\label{emax}
\end{figure}

Furthermore, the maximum stored energy $E_{\max}/(N\Delta)$ and the optimal charging period $T_{\max}$ depending on
the coupling strength are calculated in Fig.~\ref{emax}. It is observed that $E_{\max}/(N\Delta)$ approaches
to the fully-charged value $1$ in a wide range of the repulsive coupling
strengths. As the attractive
coupling strength gets close to the critical value $\lambda_c=-1$, the
maximum energy stored gets worse and drops sharply from $1$ in Fig.~\ref{emax}(a). Meanwhile,
$T_{\max}$ for the attractive coupling strength is longer
over the noninteracting case $\lambda=0$. By contrast, the charging for strong repulsive
interacting atoms with $\lambda>0$ is faster in Fig.~\ref{emax} (b). It reveals
that the strong repulsive interactions of atoms play a positive role in the
charging, which can speed up the charging of the interacting atoms.

Since the phase transition of the attractive interacting atoms is induced by the first
non-interacting $\Delta$-term and the second coupling $g$-term of the Hamiltonian
$H_0^{I}$ in Eq. (\ref{h0interacting}).
The low energy of the first non-interacting term is responsible for charging
between the states $|g\rangle$ and $|e\rangle$ for each atom. As the attractive
coupling strength increases, the evolution is
dominated by the high-energy part of the second $g$-term. The
high-energy eigenstates can influence
the charging states of the many-body battery, which play a negative role in
the charging. However, the direct correlations between the phase transition
and the charging remains unclear due to the complicated evolution in the
interacting atoms.

\section{Conclusion}

In this work, we introduce the harmonic driving field as the energy charger
for the quantum battery, which consists of $N$ two-level atoms. By contrast
to previous studies with a static charging field, the quantum battery of
noninteracting atoms can be fully charged by choosing an optimal driving
frequency. After charging process, each of two-level atoms is finally in the
upper state. Involving the intrinsic interactions between atoms, it is
clearly seen two important effects: (i) the repulsive interactions in large $%
N$ atoms can enhance the fully charging with shorter charging period,
yielding an advantage in charging over the noninteracting atoms; (ii) for
the attractive interactions case, the quantum phase transition with
degenerate energies plays a negative effect in the energy storage in this
quantum battery. The maximum stored energy of the QB in the degenerate phase
drops sharply from the fully-charging value with longer charging period.

In terms of outlook, the two-level quantum battery we discuss here could be
a physical realizable scheme. Recently, experimental efforts have been
devoted to quantum simulations of an array of two-level systems, such as
with a solid-state platform~\cite{Ferraro2018,pforn16}, trapped ions~\cite%
{kim18} and cold atoms~\cite{baumann10}, which could be considered as
quantum battery. When charging resources such as Raman laser beams with
modulated frequency are driven onto such two-level systems, the charging of
the quantum battery with a harmonic field could be implemented
realistically. The kind of quantum battery facilitates us to exploit the
contributions of the quantum correlations of many-body systems for charging
process, especially the effects of quantum phase transitions.

\acknowledgments
This work was supported by National Natural Science Foundation of China
(Grant No. 11847301) and by the Fundamental Research Funds for the
Central Universities of China (Grant No. 2019CDXYWL0029 and No. 2019CDJDWL0005).
F. L. Bin acknowledges the support from the National Natural Science Foundation of China
(Grant No. 11725417, No. 11575027, and No. 11475146) and NSAF (Grant No.U1730449).
X. G. Wang acknowledges supports from
the National Natural Science Foundation of China (Grant No. 11875231)
and the National Key Research and Development Program of China (Contracts No.
2017YFA0304202 and No. 2017YFA0205700).

\end{document}